\documentclass{article}

\PassOptionsToPackage{compress}{natbib}

\usepackage[preprint]{neurips_2025}

\usepackage[utf8]{inputenc} 
\usepackage[T1]{fontenc}    
\usepackage{hyperref}       
\usepackage{url}            
\usepackage{booktabs}       
\usepackage{makecell}

\usepackage{amsfonts}       
\usepackage{nicefrac}       
\usepackage{microtype}      
\usepackage{xcolor}         

\usepackage{amsmath} 
\usepackage{amssymb} 
\usepackage{amsthm} 
\usepackage{mathtools} 

\usepackage{subcaption}
\usepackage{float}

\usepackage{titlesec} 

\newcommand{\norm}[1]{\left\lVert#1\right\rVert}

\title{Density, asymmetry and citation dynamics in scientific literature}

\author{
    Nathaniel Imel \\
    New York University \\
    \texttt{n.imel@nyu.edu} \\
    \And
    Zachary Hafen \\
    Northwestern University \\
    \texttt{zachary.h.hafen@gmail.com} \\
}

\begin{document}

\maketitle

\begin{abstract}
    Scientific behavior is often characterized by a tension between building upon established knowledge and introducing novel ideas. Here, we investigate whether this tension is reflected in the relationship between the similarity of a scientific paper to previous research and its eventual citation rate. To operationalize similarity to previous research, we introduce two complementary metrics to characterize the local geometry of a publication's semantic neighborhood: (1) \emph{density} ($\rho$), defined as the ratio between a fixed number of previously-published papers and the minimum distance enclosing those papers in a semantic embedding space, and (2) asymmetry ($\alpha$), defined as the average directional difference between a paper and its nearest neighbors. We tested the predictive relationship between these two metrics and its subsequent citation rate using a Bayesian hierarchical regression approach, surveying $\sim 53,000$ publications across nine academic disciplines and five different document embeddings. While the individual effects of $\rho$ on citation count are small and variable, incorporating density-based predictors consistently improves out-of-sample prediction when added to baseline models. These results suggest that the density of a paper's surrounding scientific literature may carry modest but informative signals about its eventual impact. Meanwhile, we find no evidence that publication asymmetry improves model predictions of citation rates. Our work provides a scalable framework for linking document embeddings to scientometric outcomes and highlights new questions regarding the role that semantic similarity plays in shaping the dynamics of scientific reward.
\end{abstract}

\section{Introduction}

The dynamics of risk, reward and innovation has been among the most influential topics in the study of science \citep{kuhn1959essential, popper1962conjectures, lakatosFalsificationMethodologyScientific1970, fosterTraditionInnovationScientists2015, uzzi2013atypical, bourdieuSpecificityScientificField1975, fortunatoScienceScience2018}. Recent advances in natural language processing-- in particular, distributed representations of word meanings via text embeddings -- have yielded powerful tools for domain-general, efficient, and increasingly rich representations of the semantic content of scientific documents. This technology, along with neural language models, has allowed researchers in machine learning to build useful artifacts for accelerating scientific discovery \citep{wangScientificDiscoveryAge2023, krennForecastingFutureArtificial2023, zhang2024comprehensivesurveyscientificlarge, zhengLargeLanguageModels2025, ai4science2023impactlargelanguagemodels}.

It remains less clear how these tools can be used for \emph{explanation} of trends in scientific behavior. Notably, historians and sociologists of science have long argued \citep{kuhn1959essential, bourdieuSpecificityScientificField1975} that science is shaped by an `essential tension' between tradition and innovation, such that individual scientists and communities face intrinsic, often paradoxical trade-offs between producing incremental, more-traditional research and advancing novel, revolutionary ideas. Although big data and modern computational frameworks have made it possible to investigate these relationships empirically \citep{uzzi2013atypical, fosterTraditionInnovationScientists2015, fortunatoScienceScience2018}, there remain many unresolved questions about the underlying social dynamics of innovation in science. Among these, one may ask: what is the relationship between a paper's position within the existing body of knowledge and its subsequent impact?

In this paper, we investigate one measureable version of this relationship: how the similarity of a paper to previously existing research correlates with its eventual citation rate. We introduce two complementary metrics for quantifying a paper's similarity to previous research, and survey $\sim 53,000$ publications in nine academic disciplines and five different methods for obtaining document embeddings. Bayesian statistical modeling reveals that while the direct effects of these similarity-based metrics on citation count are small and uncertain, some consistently improve predictive performance when incorporated alongside established covariates. These findings suggest that a paper’s semantic proximity to prior literature encodes meaningful-- if subtle-- information about the attention it eventually receives. Altogether, our work formalizes a general framework for studying how semantic positioning within a literature landscape relates to reward, providing both tools and preliminary evidence for further empirical inquiry into the structure and dynamics of scientific innovation.

\section{Do papers that are more similar to existing papers attract more citations?}

The nature of scientific communities have long been described in terms of shared theoretical goals, methodological approaches, and epistemic attitudes \citep{kuhn1962structure, kuhn1959essential, popper1962conjectures, lakatosFalsificationMethodologyScientific1970, bourdieuSpecificityScientificField1975}. Importantly, many of these defining characteristics are linguistically encoded within \emph{documents}-- textbooks, reports, grants, and peer-reviewed articles-- that serve to transmit scientific knowledge within these communities. This shared linguistic content reflects a degree of intellectual coherence of a literature. This can be concretely modeled as a neighborhood of publications within a high-dimensional semantic space. Under this assumption, we intuitively expect that well-established or popular research areas will show a higher concentration of publications in their corresponding neighborhoods of literature. This prompts a natural question: do new contributions published into denser and more prototypical areas of literature tend to attract a higher number of citations? 

One possible outcome is that publications situated within densely populated semantic regions may, on average, accumulate higher citation rates, potentially due to the presence of a larger audience to read and cite their work. On the other hand, the generation of impactful knowledge may require introducing novel concepts, which may initially locate a publication in a less densely populated region of semantic space.
Furthermore, even within popular research areas, one might expect that new articles published farther, rather than closer, to the semantic center of these neighborhoods, may become more impactful. These potential outcomes each align with influential prior studies in the science of science  which have established that highly influential works frequently incorporate atypical or surprising conceptual combinations \citep{uzzi2013atypical, fosterTraditionInnovationScientists2015}. Such works might correspond to relatively isolated locations within the semantic landscape. 

In summary, it remains an open question whether citation rates tend to favor papers embedded within denser, more prototypical regions of the semantic landscape or those occupying sparser, more unconventional territory. Either pattern would offer valuable insight into the dynamics of scholarly attention and the diffusion of scientific ideas.

This motivation, while intuitive, leaves important notions undefined. Specifically, what constitutes the conceptual space of scientific documents, and how can we quantify the population density within a defined semantic neighborhood? To our knowledge, no prior research has leveraged domain-general semantic representations of scientific documents to directly investigate the relationship between a paper's local population density within this conceptual space and its subsequent citation impact. In this paper, we address this gap, using a variety of document embeddings to operationalize the conceptual space of scientific literature and to develop a measure for the population density surrounding individual publications.

\subsection{Related work}

Analyses of the dynamics of credit and innovation in science have long been a central focus of scholarly inquiry. \cite{fosterTraditionInnovationScientists2015} examined an ``essential tension" between tradition and innovation in biomedicine by analyzing the evolution of chemical knowledge in abstracts, revealing the trade-offs between risk and reward in different research strategies. Their approach, grounded in the extraction of fine-grained, domain-specific details of chemical compounds, contrasts with our use of more general text embeddings of documents. Before this, \cite{uzzi2013atypical} investigated novelty and scientific impact by analyzing citation patterns, finding that research with the highest impact contained both atypical and conventional combinations of prior work. Their focus on co-occurrences of references in bibliographies as a measure of novelty differs from our investigation, which focuses on the relationships between the semantic content of language in scientific publications.

More recently, the increasing availability of large text corpora and advances in natural language processing have enabled new approaches to studying scientific dynamics through the lens of semantic content. For example, \cite{yinIdentifyNovelElements2023} explored the concept of novelty by showing a relationship between expert judgments and the semantic dissimilarity of paper embeddings from their field's average. Similarly,  \cite{linNewDirectionsScience2022} demonstrated the predictive power of semantic distance, albeit at the keyword level, for citation impact. Beyond predicting impact, semantic embeddings have also been used to model broader trends in scientific behavior, as shown by \cite{liuChainingTemporalDynamics2022}, who tracked the evolution of research interests through shifts in paper embeddings. Building on this growing body of work that leverages semantic representations to understand various facets of science, our study introduces an intuitive and scalable metric of local literature density based on document embeddings to directly examine its relationship with citation rates across a large and diverse dataset.

\section{Two publication metrics based on similarity to previous research}

To understand how a scientific paper's relationship to prior research might influence its future impact, we focus on two, complementary operationalizations of semantic similarity of publications within the conceptual space of scientific literature: \textbf{density} and \textbf{asymmetry}. 

In order to do this, we first represent the semantic content of scientific documents as vectors in a high-dimensional space, referred to as a \emph{document embedding space}. Generally, text embedding techniques proceed by constructing numerical representations for words or documents by analyzing statistical patterns in language usage within large text corpora \citep{jurafsky2025speech}. These representations aim to capture the meaning and relationships between linguistic units. One simple approach is to create ``bag-of-words" vectors, which are essentially feature vectors based on word frequencies. More modern techniques typically involve deriving vectors from the hidden representations of neural network models (including large language models) trained to learn co-occurrence statistics of linguistic units. These latter techniques have produced rich and context-sensitive vector representations \citep{petersDeepContextualizedWord2018}.

Given a document embedding space, we find the $k$ papers that are most similar to a given paper – its $k$ nearest neighbors \citep{fix1951discriminatory, coverNearestNeighborPattern1967}. The size or extent of this neighborhood is then determined by the `distance' needed to encompass these closest papers in semantic space. As is standard practice, we use cosine distance as our measure of semantic dissimilarity. When these vectors are normalized to have a length of one, the cosine distance simplifies to the angle between them.

\subsection{Density}
\begin{figure*}[]
    \centering
    \includegraphics[width=0.8\linewidth]{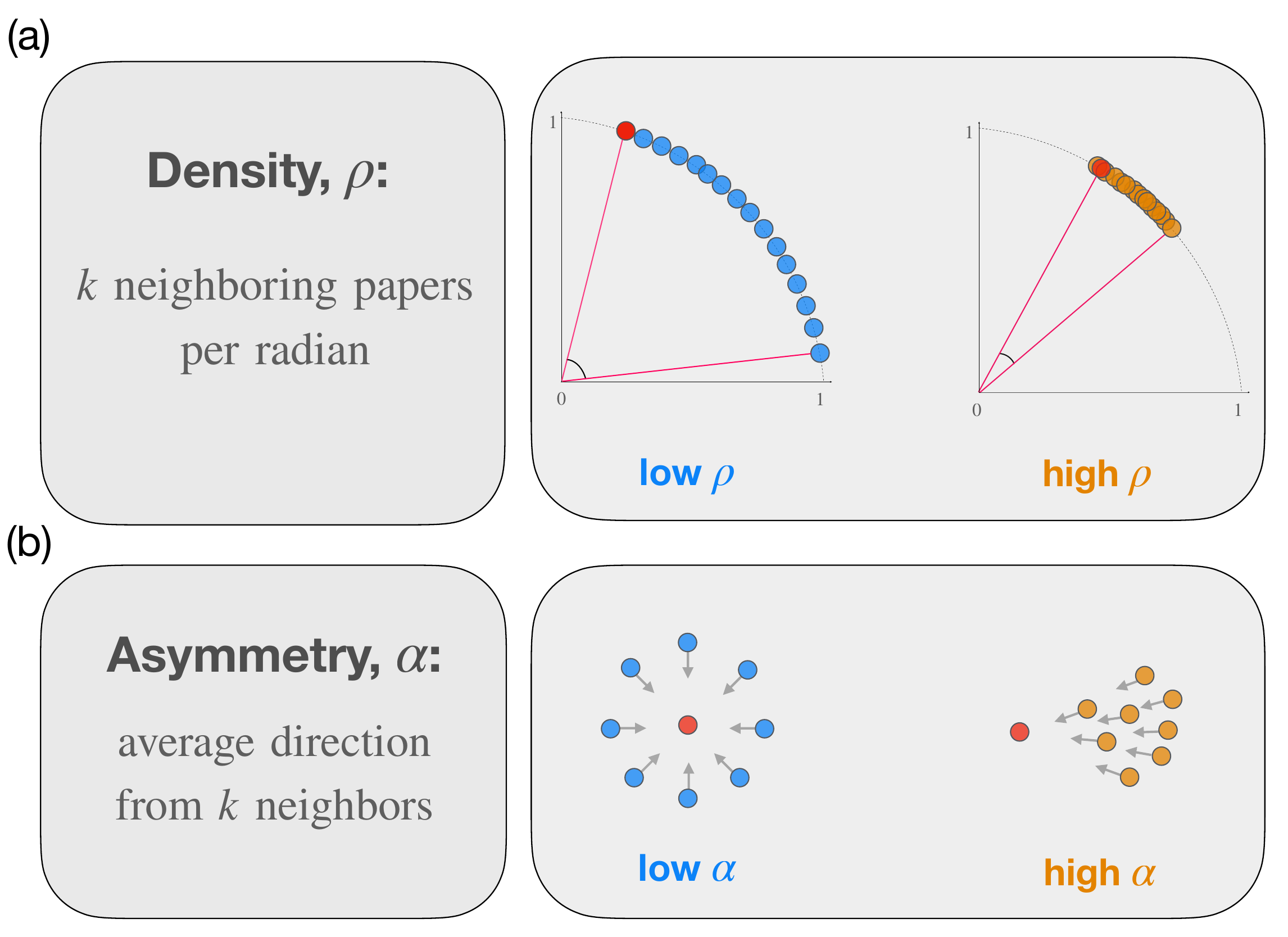}
    \caption{ Two document similarity-based publication metrics. \textbf{(a)}: Our ``density" metric, $\rho$, which is defined in Equation \ref{eq:density} and captures for a given publication $p_i$ the ratio of $k$ nearest neighboring publications to the arclength enclosing them in document embedding space. \textbf{(b)} Our ``asymmetry" metric, $\alpha$, which is defined in Equation \ref{eq:asymmetry} and captures for a given publication $p_i$ the average directional difference between it and its $k$ semantic neighbors. }
    \label{fig:approach_figure}
\end{figure*}
Formally, for a publication $p_i$ represented by its embedding $\mathbf{p}_i$ in an $n$-dimensional space, we define its density $\rho(p_i)$ based on its $k$ nearest \emph{previous} neighbors. We calculate this density as the ratio of $k$ to the angle required to enclose these neighbors (analogous to a kernel size; \cite{terrellVariableKernelDensity1992}). A publication's density, $\rho$, is given by:
\begin{equation}
    \label{eq:density}
    \rho(p_i) = \frac{k}{ \arccos ~ \mathbf{p}_i^{\top} \mathbf{p}_k },
\end{equation}
where $\mathbf{p}_k$ is the embedding of $p_i$'s $k^{th}$ nearest previous neighbor. When all document embeddings are normalized to unit length, the inner product $\mathbf{p}_i^{\top} \mathbf{p}_k$ is equivalent to their cosine similarity, and $\arccos(\mathbf{p}_i^{\top} \mathbf{p}_k)$ yields the angle between them. Thus, $\rho(\mathbf{p_i})$ quantifies the number of nearby previous publications per radian in the semantic space.

Traditional physical density scales as $\propto 1/{\rm{distance}}^d$, where $d$ is the number of dimensions. We adopt $\rho$, as a simpler measure, for three reasons. First, our goal is to capture meaningful social dynamics, not to strictly mirror physical analogies. Second, defining density as $\propto 1/ (\mathbf{p}_i^{\top} \mathbf{p}_k) ^d$ would depend on the number of embedding dimensions $d$, which is sensitive to model architecture and not inherently meaningful. Third, such formulations can lead to numerical underflow when $d$ is large and $ \arccos ~ \mathbf{p}_i^{\top} \mathbf{p}_k < 1$, as is the case here. Although $\rho$ is not a true volumetric density, it remains a principled proxy: it is a monotonic transformation of $k$-nearest neighborhood density that preserves rank information and can be interpreted as a surface density on the unit hypersphere. 

\subsection{Asymmetry metric}

Our asymmetry metric captures a qualitatively distinct aspect of semantic similarity: how much a publication lies at the `edge' of its local neighborhood. It corresponds to the magnitude of the net direction formed by the $k$-nearest neighboring publication vectors. Formally, the asymmetry of a publication $\mathbf{p_i}$ is:
\begin{equation}
    \label{eq:asymmetry}
    \alpha(\mathbf{p_i}) = 
     \frac{1}{k} \norm{ ~ \sum_{j=1}^{k} \frac{\mathbf{p_i }- \mathbf{p_j}}{| \mathbf{p_i} - \mathbf{p_j} |} ~ },
\end{equation}
where $\norm{\cdot}$ is the Euclidean norm. This metric also takes inspiration from physics, in the following sense: if publications are thought of as particles subject to pairwise forces, then asymmetry is similar to net force, which is $0$ if the forces contributed by surrounding particles on $p_i$ cancel. If a publication has $0$ asymmetry, it can be thought of as highly prototypical, being at the semantic center of its neighborhood. One key difference between asymmetry and net force is that we remove any magnitude information by using only unit vectors. This serves to decouple the definition of asymmetry from the definition of density. e.g. it is possible for a publication to have high $\rho$, but either low or high asymmetry, as depicted in Figure \ref{fig:approach_figure}b.

\section{Methods}

\subsection{Citation rates}
\label{sec:citations}
Given the two metrics we introduced for quantifying a publication's similarity to previous literature-- density and asymmetry -- it is natural to ask how they might relate to their eventual scientific impact. Quantifying the impact of individual papers is generally challenging, and there is no universally accepted metric. Here, we follow others \cite{fortunatoScienceScience2018, fosterTraditionInnovationScientists2015} in assuming that \emph{citations} reflect fundamental currencies of scientific recognition, and that they can serve as an indicator (albeit an imperfect one) of the degree of attention or interest in a topic for a scientific community. 

We consider the mean \textbf{citation rate} of a publication to be the total number of citations it has received until present, divided by the number of years since its date of publication in Semantic Scholar. Note that citations are events that only occur after an article was published. In order to quantify a publication's similarity to \emph{previous} publications, and also to ensure that any emergent trends in our metrics with citations do not trivially result as a matter of definition, we exclude all subsequently appearing publications from analysis when calculating our semantic metrics for each publication.

In our experiments, we measure $\rho$, and subsequent citations per year, for each paper in a sample of roughly $53,000$ articles from Semantic Scholar \cite{Kinney2023TheSS} across $9$ disciplines between 2000 and 2020, repeating this measurement with $5$ kinds of embeddings. 
We release our data and code at \url{https://github.com/nathimel/citesim}, in addition to a pip-installable software library designed to support replication and extension of this style of analysis for different publication metrics, embeddings, and bibliographic APIs: \url{https://pypi.org/project/sciterra/}.

\paragraph{Data} Our dataset is a sample from the Semantic Scholar (S2) Academic Graph \cite{Kinney2023TheSS} via the free S2 API. For each field, we chose a random initial publication `center', and iteratively retrieved from Semantic Scholar (S2) at least $30,000$ papers from the citation network in order of decreasing cosine similarity of the SciBERT embedding from this center. We required that each new publication added have (i) an abstract (ii) a publication date, (iii) were associated with the target field of study. This criterion was satisfied by only 10\% of papers retrieved from S2. We further restrict our analysis to papers from this sample that were (i) published between 2000 and 2020 (which mitigates bias in the dataset towards highly influential older publications with sparser neighborhoods) and (ii) did not change neighborhood composition over the period of updates during which $N=1000$ publications were added to the dataset. Selecting the size of this neighborhood -- the number $k$ of nearest neighbors that must not change after multiple expansions -- is a free parameter. There is a trade-off in selecting its value: with too few neighbors, density estimates fail to be informative; requiring each neighborhood to include all data points results in 0 total converged publications for analysis. We selected $k=16$, which yielded yielded $53,080$ data points across all fields (ranging between $4,114$ and $7556$ data points per field) for analysis. For further discussion of this technique, we refer readers to \cite{imelCitationSimilarityRelationshipsAstrophysics2023}.

\begin{figure}
    \centering
    \includegraphics[width=0.99\linewidth]{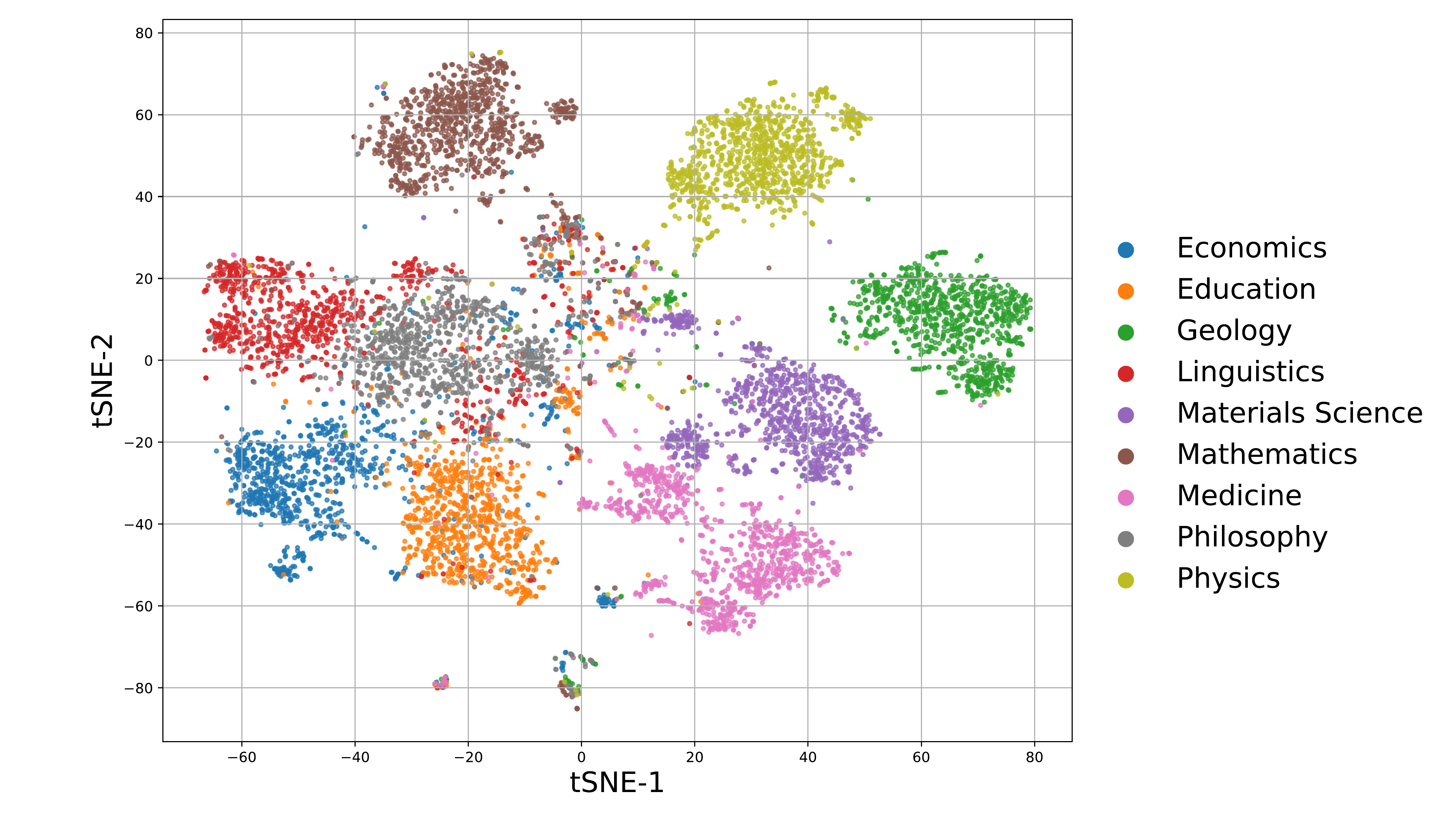}
    \caption{Visualization of a sample of $9000$ scientific publications in semantic space by field using document embeddings obtained from SciBERT \citep{beltagy2019scibert}. Each point represents a paper using a document embedding of its abstract that is projected into 2 dimensions via t-SNE dimension reduction.}
    \label{fig:tsne_scibert}
\end{figure}

\paragraph{Document embeddings}
We computed measurements of $\rho$ using five different document embedding models: SciBERT \cite{beltagy2019scibert},  SBERT \cite{Reimers2019SentenceBERTSE}, GPT-2 \cite{radford2019language}, averaged skip-gram Word2Vec word embeddings \cite{Mikolov2013EfficientEO}, and a simple bag-of-words (BOW) approach. To obtain embeddings from SciBERT for each abstract, we obtained a $768$ dimensional vector by extracting the final hidden state of the $\texttt{[CLS]}$ token from SciBERT given the text of the paper's abstract (but not SPECTER embeddings, which contain information about the citation graph \cite{cohan-etal-2020-specter}). For GPT-2, we use the final token of the final hidden state, which also results in a $768$ dimensional vector to represent each paper's abstract. For SBERT, we follow \cite{liuChainingTemporalDynamics2022} in treating each abstract as a sentence and use the resulting embeddings. For our Word2Vec abstract embeddings, we obtain $300$ dimensional word vectors for each word in the abstract and take their average. This was done by training a separate skipgram Word2Vec model with default parameters for each field, using all of the text of the abstracts (roughly $30,000$) for that field. For our bag-of-words embeddings, we use for each field the vocabulary from that field's respective Word2Vec model and use word counts for feature values, resulting in embedding dimensions of roughly $17,000$. While these word vectors do not capture as rich notions of conceptual similarity as embeddings from more recent language models, they are perhaps more interpretable, corresponding to an approximate measure of text overlap. As a sanity check, we verified that field labels could be reliably predicted from the embeddings using a simple linear classifier, confirming that the representations captured meaningful semantic structure. A visualization of a sample of SciBERT embeddings projected down to $2$ dimensions using t-SNE \citep{van2008tsne} and colored by field is depicted in Figure \ref{fig:tsne_scibert}. For visualizations of the other embeddings used in our analysis, and more details of this sanity check, see Appendix \ref{app:tsne}.

\paragraph{Bayesian statistical modeling}

To quantify the strength of the correlations between our semantic metrics (density and asymmetry; $\rho$ and $\alpha$ respectively) and citation impact, alongside known predictors, we employed Bayesian statistical modeling using STAN~\citep{stan2024}. Specifically, we fit hierarchical linear regression models to predict a publication's citation rate, accounting for age and reference count, which are established predictors \citep{mammolaImpactReferenceList2021}, as well as $\rho$ and $\alpha$ derived from each of our five embedding models. Based on well-known properties of citation patterns across disciplines \citep{radicchiUniversalityCitationDistributions2008}, our models assume that the logarithm of citations per year ($\log\rm{cpy}$) follows a normal distribution: $\log\rm{cpy} \sim N(\mu, \sigma)$. We fit the mean $\mu$ as a linear combination of our input predictors:
\begin{equation}
    \label{eq:mean_prediction}
    \mu = \gamma + \sum_{l=1}^{L} \beta_{lf} x_{l}
\end{equation}
In Equation \ref{eq:mean_prediction}, $\gamma$ is the global intercept, and $x_{l}$ represents the $l$-th predictor (where $l$ ranges from $1$ to $L$, the total number of predictors such as age, reference count, and the different density and asymmetry estimates). Prior to fitting, $x_{l}$ and $\log\rm{cpy}$ are standardized to have a mean of $0$ and a standard deviation of $1$ within each field, which enables a fair comparison of the importance of different predictors. The coefficient $\beta_{lf}$ is the fitted weight associated with the $l$-th predictor for a given field of study $f$, and corresponds to the expected change in the standardized $\log\rm{cpy}$ for a one standard deviation change in that predictor within that specific field.

The hierarchical aspect of our model arises from how we model these field-specific coefficients. Rather than treating each $\beta_{lf}$ as entirely independent, we assume they are drawn from a shared normal distribution: $\beta_{lf} \sim N(\mu_{l}, \sigma_{l})$, where $\mu_{l}$ is the underlying mean across all fields and $\sigma_{l}$ is the deviation between means. This approach allows us to directly measure the relationship between coefficients in different fields and also generates more stable and robust estimates, especially when individual field-level data might be noisy.

To assess the robustness of our findings, we explored $27$ variations of this base model. These variations included: a) dropping the dependence on specific predictors (setting $\beta_{lf} = 0$ for a given $l$), b) constraining the effect of certain predictors to be consistent across all fields (setting $\beta_{lf} = \beta_l$), and c) for the density estimates, sampling the mean of the field-level coefficients ($\mu_{l}$) from another shared normal distribution. Notably, the estimated coefficients $\beta_{lf}$ remained relatively stable, varying by less than 10\% across these model variants. The performance of all model variations was evaluated on a held-out test set comprising 25\% of our data using six distinct metrics: root mean squared error, mean absolute error, mean absolute percentage error, $R^2$, and the $D^2$ pinball and absolute error scores. All six metrics consistently identified the same best-fit model and yielded the same performance ranking across the other model variations, providing confidence in our model selection and evaluation.

\section{Results}

\begin{figure*}[h!]
    \centering
    \includegraphics[width=0.99\textwidth]{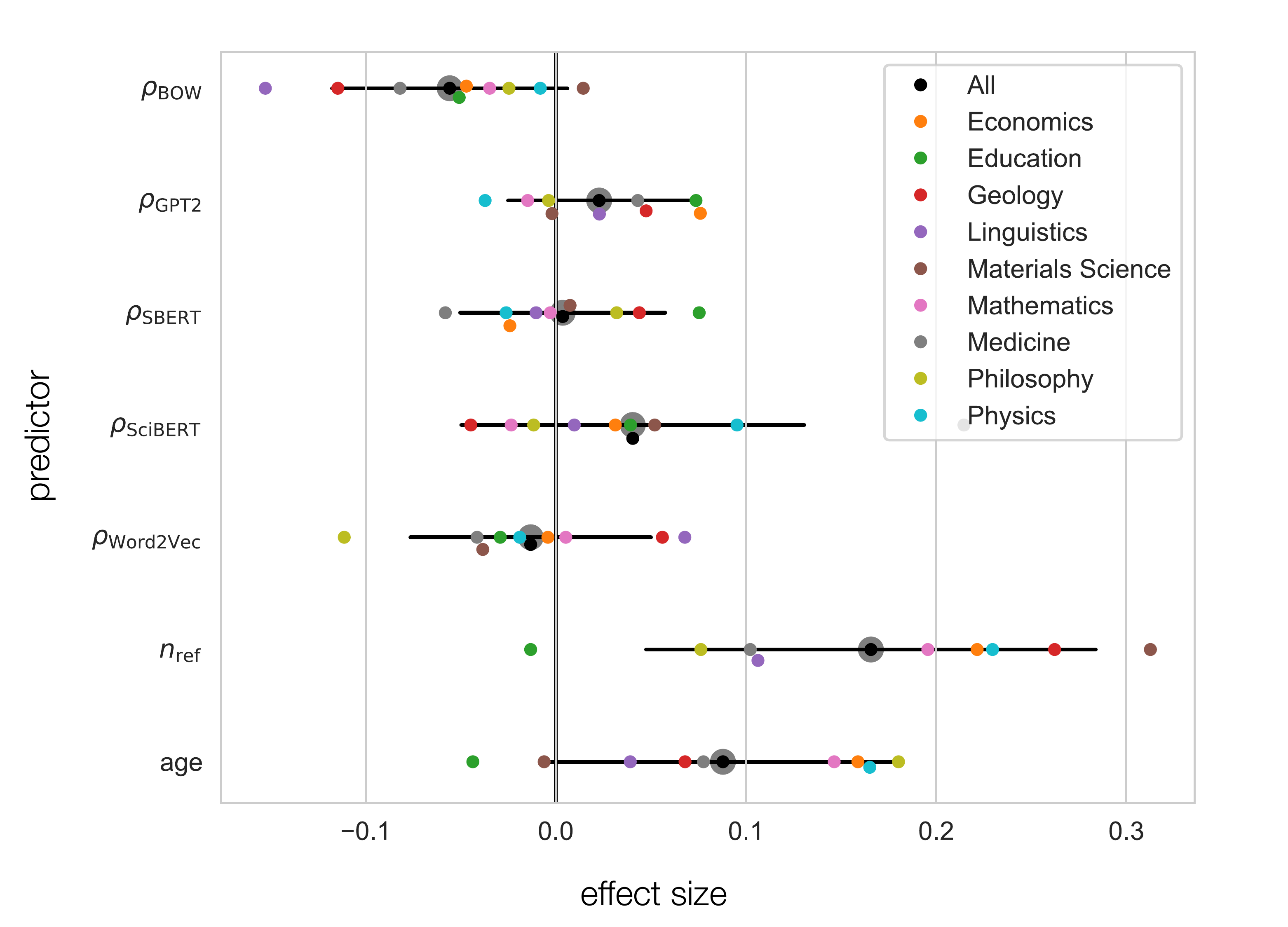}
    \caption{
    Estimated effect sizes ($\beta_{lf}$) from our best-performing hierarchical linear regression model, predicting the logarithm of citations per year (scaled within each field) based on reference count, publication age, and our density metric ($\rho$) calculated for each of five different embedding models, where $l$ denotes the predictor and $f$ denotes the academic field. Including asymmetry ($\alpha$) estimates as predictors did not improve model performance (see Appendix \ref{app:models}). For each predictor $l$, the field-specific effect sizes $\beta_{lf}$ (colored points) were modeled as being drawn from a shared normal distribution with a mean $\mu_{l}$ (large black points) and standard deviation $\sigma_{l}$ (solid black lines encompassing). Across fields, the effect size is typically negative for $\rho_{\mathrm{BOW}}$; consistent with zero (gray vertical line) for $\rho_{\mathrm{GPT2}}$, $\rho_{\mathrm{SBERT}}$, and $\rho_{\mathrm{Word2Vec}}$; and positive for reference count, publication age, and possibly $\rho_{\mathrm{SciBERT}}$.
    }
    \label{fig:beta_coeffs}
\end{figure*}

Figure \ref{fig:beta_coeffs} visualizes the estimated effect sizes ($\beta_{lf}$) for our best-fitting hierarchical linear regression model. This model allowed the effect of each predictor -- the five embedding-derived density estimates ($\rho$), the publication's reference count, and its age -- to vary across the nine academic fields in our dataset. This model achieved modest but consistent increase in out-of-sample performance across evaluation metrics as compared to a model that included field-specific reference count and age but did not include any $\rho$ estimates (Table \ref{tab:models}). 

\newlength{\oldtabcolsep}
\setlength{\oldtabcolsep}{\tabcolsep}
\setlength{\tabcolsep}{5pt} 

\begin{table}[h]
\centering
\small
\begin{tabular}{l@{\hskip 6pt}ccccc}
\toprule
  & 
Base / Mean & 
\makecell{Per-field\\$(n_{\rm ref}, t)$} & 
\makecell{Per-field, per-model\\$(\rho, n_{\rm ref}, t)$} & 
\makecell{Per-field, per-model\\$(\alpha, n_{\rm ref}, t)$} & 
\makecell{Per-field, per-model\\$(\rho, \alpha, n_{\rm ref}, t)$} \\
\midrule
$1 - \text{RMSE}$ & 0.009 & 0.024 & \textbf{0.037} & 0.025 & 0.028 \\
\bottomrule\\
\end{tabular}
\caption{Model performance measured by $1 - $ root mean squared error, where higher values indicate better predictions. Including $\rho$ (density) yields improvement, while additionally including $\alpha$ (asymmetry) lowers performance. See Appendix \ref{app:models} for further details. } 
\label{tab:models}
\end{table}

This suggests that including information about a paper's semantic neighborhood density can explain some of the variance in its subsequent citation rate. Importantly, as Figure \ref{fig:beta_coeffs} shows, there is wide variability in the magnitude of this trend across different disciplines, and different document embeddings likely contribute different aspects of this information. Notably, the density metrics derived from all embedding models had credible ($\pm \sigma$) intervals of the effect sizes that were consistent with zero. This means the variables are mildly informative, but not strong enough to stand out on their own, especially under Bayesian shrinkage.

Among the density estimates, $\rho$ derived from bag-of-words (BOW) embeddings showed the highest magnitude effect ($\beta_{\rm BOW} \approx -0.05 \pm 0.06$), suggesting a modest negative association between BOW-derived density and citation rate. The density estimates with the second highest magnitude were those derived from SciBERT ($\beta_{\rm SciBERT} \approx -0.04 \pm 0.09$). As anticipated, the number of references cited ($\beta_{\rm ref} \approx 0.14 \pm 0.13$) and the age of the publication ($\beta_{\rm age} \approx 0.08 \pm 0.09$) exhibited the largest effect sizes on the logarithm of citations per year. 

Surprisingly, including asymmetry estimates as predictors did not increase model performance (compared to a model that included field-specific reference count and age but did not include any $\alpha$ estimates), and for some evaluation metrics, even decreased model performance. Even when adding asymmetry estimates as predictors to our main reported model, this did not significantly, consistently improve model performance (for further details and visualizations on these comparisons, see Appendix \ref{app:models}). 

Overall, while the estimated effects of $\rho$ across embeddings and fields were individually small and uncertain, their aggregate inclusion nevertheless improved model performance across all evaluation metrics. This suggests that our metric for quantifying the local semantic neighborhood density of publications
captures weak but systematic structure in citation patterns. In contrast, we did not observe consistent evidence that asymmetry estimates improved predictive performance. In the following section, we discuss the implications of these results and outline some directions for future work.

\section{Discussion}

This study set out to investigate whether the position of a scientific publication within a semantic landscape can help explain variation in its subsequent citation rate, a question relevant to long-standing debates regarding the tension between tradition and innovation \citep{kuhn1959essential, fosterTraditionInnovationScientists2015}. To implement this analysis, we introduced a scalable computational framework for quantifying the local density of a region of literature based on document embeddings and measuring density and asymmetry alongside other scientometric variables. We conducted a careful analysis of these variables using a Bayesian hierarchical regression approach across a large-scale dataset of publications across different disciplines and document embeddings. This allowed us to conservatively assess the relationship between literature density and citation impact while accounting for field-specific variations and established predictors. 

We found that density-based predictors, despite their small and uncertain individual effects, consistently improved model performance when included alongside baseline covariates. In particular, although our individual posterior effect sizes are close to null, $\rho$-based predictors can still improve out-of-sample prediction. Notably, asymmetry-based predictors did not further improve performance, suggesting that different facets of semantic similarity vary in their predictive utility and that density-based measures may better align with the structure relevant to citation behavior. Taken together, our results show that although the similarity of a paper to the existing literature may have only a subtle direct effect on citation rates, it nonetheless contributes meaningful information that helps explain citation patterns when considered together with other factors.

There are several important limitations to our study that should be considered when interpreting these results. First, we did not validate our semantic similarity metric with human or expert annotations. While obtaining consistent judgments of the similarity of scientific documents may be practically challenging and difficult to scale, it may also represent the most principled way forward to validate computational metrics derived from the semantic representation of documents and is therefore an important avenue for future research. If feasible, this could open new avenues to explore the robustness of semantic similarity as a predictor of citation rates.  Second, while our sample of approximately $53,000$ publications is substantial, it is smaller than the millions of publications often analyzed in contemporary scientometric studies. This sample size was chosen based on the convenience of accessing open-source materials and a commitment to utilizing the metadata available through Semantic Scholar. A larger-scale analysis leveraging other data sources would be a valuable next step, and our publicly available analysis code (as a documented and pip-installable software package, \url{https://pypi.org/project/sciterra/}) should facilitate such replication. Third, due to data constraints, we were unable to control for potentially influential variables such as the h-index of affiliated authors or the impact factor of the publication journal. Future research controlling for these factors could provide a more comprehensive understanding of the role of the metrics considered here on citation impact. Finally, our finding of a largely null predictive relationship for any individual estimate of density and asymmetry could be evidence that further refinement of how we conceptualize and measure a publication's local semantic environment is needed. Future research should investigate alternative approaches to operationalizing density, asymmetry and neighborhood. 

Altogether, our work lays the groundwork for a computational approach to understanding how semantic similarity may shape scientific recognition, and offers several promising avenues for future research.

\bibliographystyle{unsrtnat}


\appendix

\section{Model comparisons}
\label{app:models}

\begin{figure}[!h]
    \centering
    \includegraphics[width=0.99\linewidth]{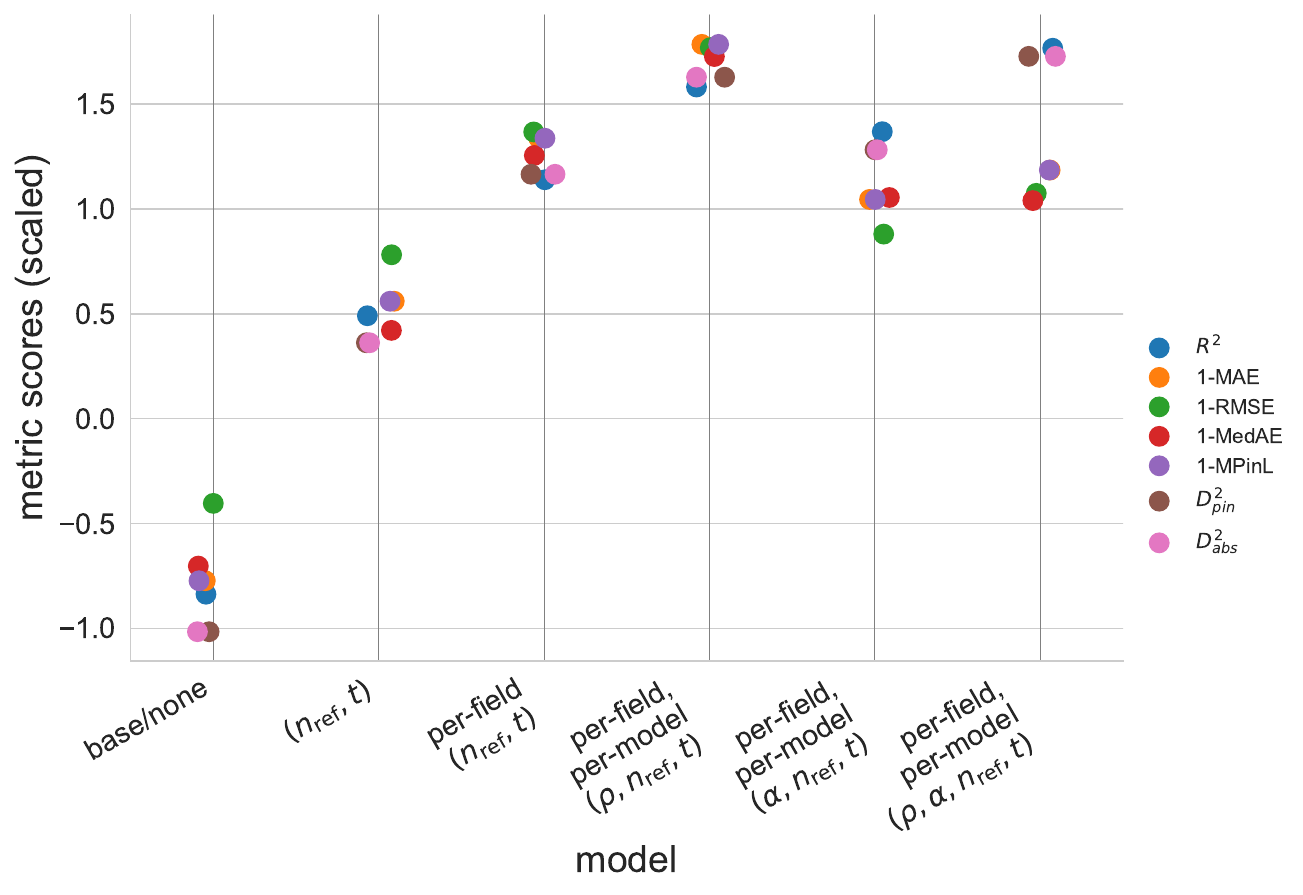}
    \caption{ Comparison of different Bayesian hierarchical models for predicting citation rates ($\log\rm{cpy}$). The $y$-axis marks the scores for different evaluation metrics, scaled within the models considered for rank ordering.
    Different models are marked on the $x$-axis: base/none indicates a baseline model that predicts the mean; $(n_{\mathrm{ref}}, t)$ is a model with coefficients for reference count and publication age; per-field $(n_{\mathrm{ref}}, t)$ is a model that includes hierarchical coefficients for reference count and publication age, each drawn from a shared normal distribution with a fixed mean; per-field, per-model $(\rho, n_{\mathrm{ref}}, t)$ similarly includes hierarchical coefficients for $\rho$ based on embedding method and field, and corresponds to our reported model in the main text. The remaining two models include hierarchical coefficients for asymmetry $\alpha$ estimates.
    }
    \label{fig:models}
\end{figure}

\begin{figure}[!h]
    \centering
    \includegraphics[width=\linewidth]{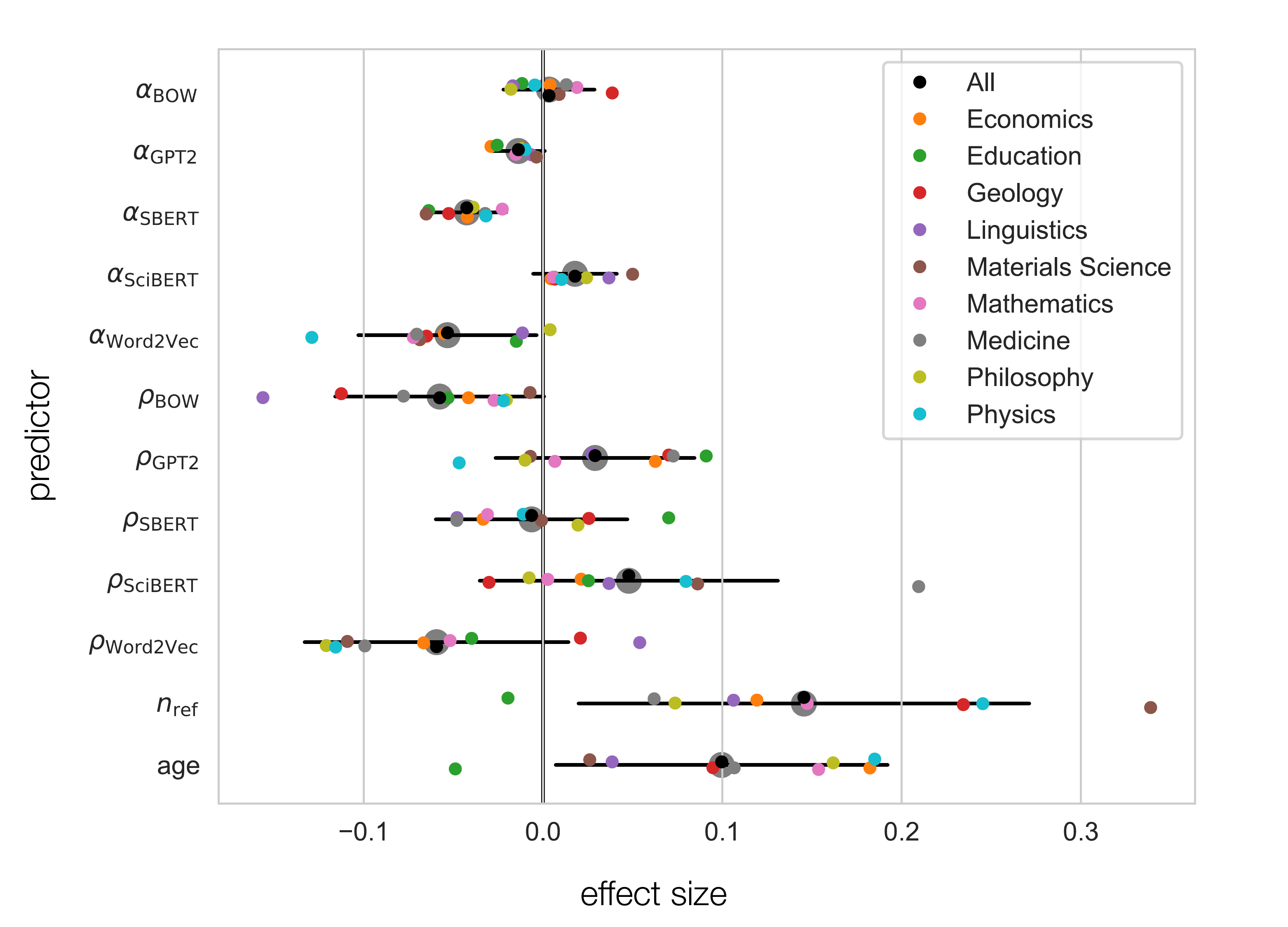}
    \caption{ 
    Estimated effect sizes ($\beta_{lf}$) from a hierarchical linear regression model that includes both density ($\rho$) and asymmetry ($\alpha$) predictors for each of five embedding models, in addition to reference count and publication age. This model extends the main analysis reported in the text by including asymmetry terms. As in the main model, field-specific effects $\beta_{lf}$ (colored points) are drawn from shared distributions with mean $\mu_l$ (large black points) and standard deviation $\sigma_l$ (black lines). While asymmetry estimates show relatively stable posterior distributions, their inclusion does not substantially alter the overall pattern of effects or improve model performance.    
    }
    \label{fig:effect_sizes_extended}
\end{figure}

Here we show the metrics across different hierarchical models we tested. It can be seen in Figure \ref{fig:models} that adding asymmetry estimates as predictors did not significantly and consistently improve performance on evaluation metrics. Furthermore, a model including coefficients for all density estimates and all asymmetry estimates did not significantly improve upon a model with coefficients for all density estimates, which is the main model reported in our results. For completeness, we include an analogous visualization of the effect sizes under this latter, more complicated model in Figure \ref{fig:effect_sizes_extended}. While the asymmetry posterior effect sizes are more consistent across fields (lower variance) than density effect sizes, we suspect this is because they do not contribute significantly to the model's performance.  


\newpage

\clearpage

\section{Clustering of embeddings across fields}
\label{app:tsne}

\begin{figure}[!h]
    \centering
    \includegraphics[width=0.99\linewidth]{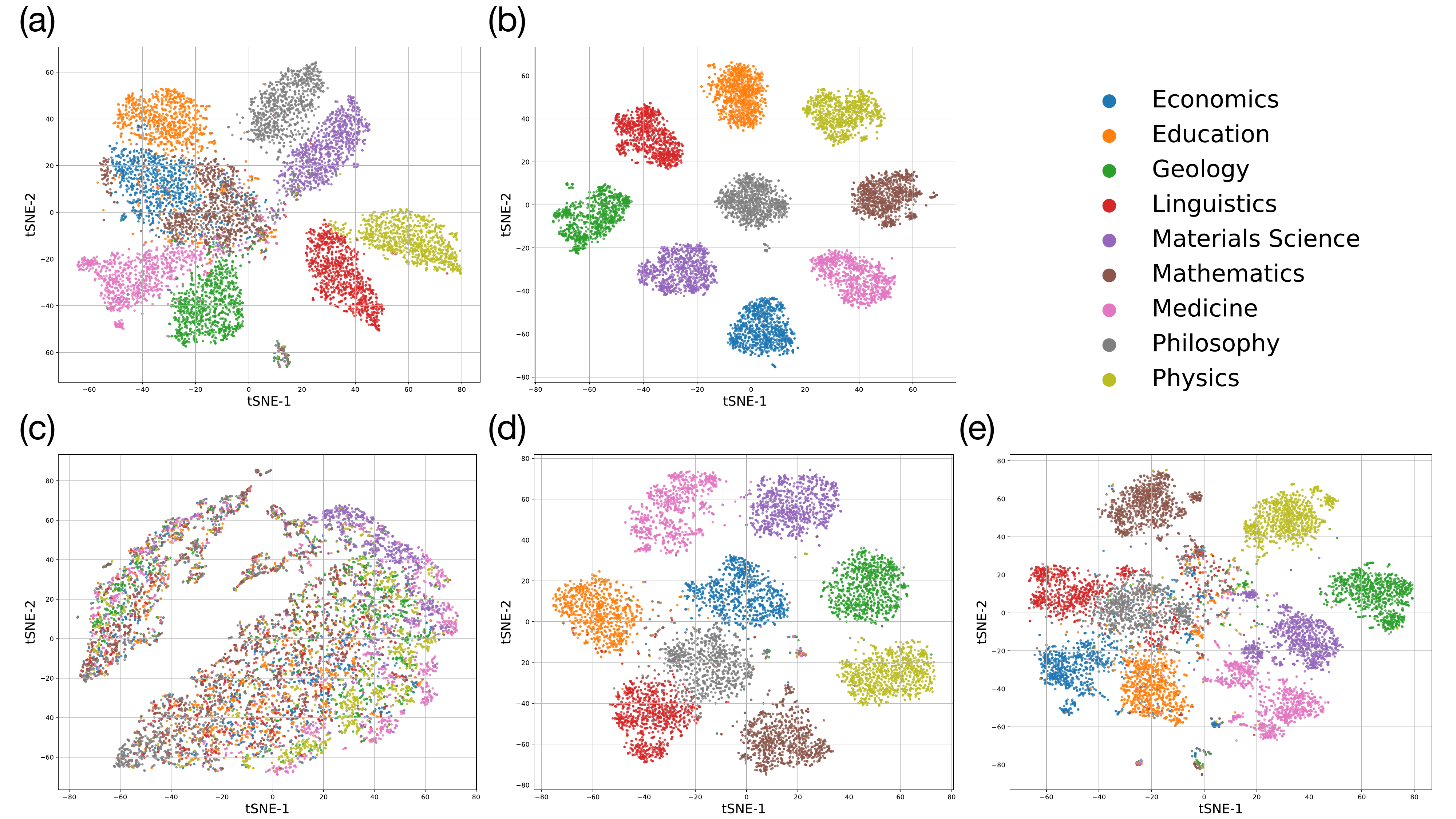}
    \caption{ Visualization of scientific publications in semantic space by field. Each point represents a paper using a document embedding of its abstract that is projected into 2 dimensions via t-SNE dimension reduction, with each panel corresponding to embeddings based on \textbf{(a)} a BOW approach \textbf{(b)} averaged Word2Vec embeddings, \textbf{(c)} GPT2, \textbf{(d)} SBERT and \textbf{(e)} SciBERT. Each panel visualizes a sample of $1000$ embeddings from each field. 
    }
    \label{fig:tsne}
\end{figure}

To qualitatively assess whether embeddings capture enough semantic information to distinguish field-specific structure, we visualized a pooled sample of embeddings from multiple fields using t-SNE dimensionality reduction \citep{van2008tsne}. Each panel in Figure~\ref{fig:tsne} shows the resulting 2D projection for a different embedding method, with colors indicating the field of origin. For computational reasons, we first reduced the dimensionality of the pooled BOW vectors -- which had an original dimensionality of nearly $100,000$ to 100 using SVD (using $\texttt{sklearn.decomposition.TruncatedSVD}$ ) before applying t-SNE. 

We also validated these patterns quantitatively by training a simple logistic regression classifier to predict the field of a paper from its embeddings. We found that classification performance was near ceiling (see Table~\ref{tab:classification}). Except for GPT2-based embeddings which also did not separate cleanly into clusters in the t-SNE 2D projection, this was true whether using the full embeddings or the 2D projections shown in Figure \ref{fig:tsne}. This analysis confirmed a basic sanity check that the different embedding methods encode sufficient signal to distinguish between academic fields.

\setlength{\tabcolsep}{\oldtabcolsep}

\begin{table}[!h]
    \centering
    \begin{tabular}{lcccccc}
        \toprule
        & BOW & Word2Vec & GPT2 & SBERT & SciBERT \\
        \midrule
        Full embeddings     &  0.98    &  1.0      &  0.94     &   0.98     &   0.95      \\
        Reduced (2D)        &  0.85    &  1.0      &  0.25     &   0.96     &   0.85      \\
        \bottomrule\\
    \end{tabular}
    \caption{Classification accuracy of a logistic regression classifier trained to predict field labels from full and 2D-reduced embeddings (using an 80/20 train-test split). }
    \label{tab:classification}
\end{table}

\end{document}